\documentstyle [12pt,epsf] {article}

\catcode`@=12
\def\bra#1{\left\langle #1\right|}
\def\ket#1{\left| #1\right\rangle}
\def\be{\begin{equation}}
\def\ee{\end{equation}}
\def\lsim{\mathrel{\vcenter{\hbox{$<$}\nointerlineskip\hbox{$\sim$}}}}
\newcommand{\bea}{\begin{eqnarray}}
\newcommand{\eea}{\end{eqnarray}}
\newcommand{\nn}{\nonumber}
\newcommand{\Apa}{A_\parallel}
\newcommand{\Ape}{A_\perp}
\newcommand{\dab}{\delta_{ab}}
\newcommand{\dca}{\delta_{ca}}
\newcommand{\dcb}{\delta_{cb}}
\newcommand{\da}{\delta_{a}}
\newcommand{\db}{\delta_{b}}
\newcommand{\dc}{\delta_{c}}
\newcommand{\pab}{\varphi_{ab}}
\newcommand{\pcb}{\varphi_{cb}}
\newcommand{\pca}{\varphi_{ca}}
\newcommand{\pa}{\varphi_{a}}
\newcommand{\pb}{\varphi_{b}}
\newcommand{\pc}{\varphi_{c}}

\def\sss{\scriptscriptstyle}
\begin{document}
\vspace{0.5in}
\oddsidemargin -.375in
\newcount\sectionnumber
\sectionnumber=0
\def\bra#1{\left\langle #1\right|}
\def\ket#1{\left| #1\right\rangle}
\def\be{\begin{equation}}
\def\ee{\end{equation}}
\def\lsim{\mathrel{\vcenter{\hbox{$<$}\nointerlineskip\hbox{$\sim$}}}}
\thispagestyle{empty}


\def\sss{\scriptscriptstyle}
\def\barp{{\raise.35ex\hbox
{${\sss (}$}}---{\raise.35ex\hbox{${\sss )}$}}}
\def\barpd{{\raise.35ex\hbox
{${\sss (}$}}--{\raise.35ex\hbox{${\sss )}$}}}
\def\dbbarp{\hbox{$B^{0}$\kern-1.2em\raise1.5ex\hbox{\barpd}}}
\def\kbarp{\hbox{$K^{*0}$\kern-1.6em\raise1.5ex\hbox{\barpd}}}

\def\dkbarp{\hbox{$K^{*}$\kern-1.2em\raise1.5ex\hbox{\barpd}}}
\vskip0.5truecm

\def\tr{\mbox{tr}\,}
\def\Tr{\mbox{Tr}\,}
\def\dag{^\dagger}
\def\res{\mbox{Res}}
\def\re{\mbox{Re}\,}
\def\b{\bigskip}
\def\s{\smallskip}
\def\l{\hspace*{0.05cm}}
\def\esp{\hspace*{1cm}}

\def\be{\begin{equation}}
\def\ee{\end{equation}}
\def\bea{\begin{eqnarray}}
\def\eea{\end{eqnarray}}

\begin{center}

{\large \bf
\centerline{Searching for new physics in the angular 
distribution of}}
{\large \bf \centerline {\boldmath{$B_d^0 \rightarrow \phi K^{*0}$} 
 decay}}
\vspace*{1.0cm}
{ Anjan K. Giri$^1$ and  Rukmani Mohanta$^2$ } \vskip0.3cm
{\it  $^1$ Physics Department, Technion-Israel 
Institute of Technology, 32000 Haifa, Israel}\\
{\it $^2$ School of Physics, University of Hyderabad,
Hyderabad - 500046, India
} \\
\vskip0.5cm
\bigskip
(\today)
\vskip0.5cm

\begin{abstract}
Motivated by the possible discrepancy between the observed CP asymmetry and
that of the standard model expectation in the decay mode $B^0\to \phi K_S$,
we study the corresponding vector vector decay mode $B^0 \to  \phi K^{*0}$.
~In order to obtain decisive information regarding the CP violation effect,
we made the angular distribution analysis of the decay products, where 
both the outgoing vector mesons decay into two pseudoscalars. Furthermore,  
we study the possible effects of new physics using the angular distribution
observables.
\end{abstract}
\end{center}
\thispagestyle{empty}
\newpage
\baselineskip=14pt

\section{Introduction}

The study of $B$ physics provides a great opportunity and an ideal testing
ground to
obtain a deep insight into the flavor structure of the standard model (SM)
and the origin of CP violation. In view of the wide variety of decay
channels one can look for many different observables, providing
stringent tests for the consistency of the model.
The goal of the $B$ factories is not only to test the SM  picture
but also to discover the evidence of new physics (NP). Recently,
the measurement of the time dependent CP asymmetries in the $B \to \psi K_S$ 
decays has led to the confirmation of CP violation in $B$ systems. The 
observed world average of the asymmetry, i.e., $\sin 2 \beta $ 
is given by \cite{nir1}
\be
\sin(2 \beta)_{\psi K_S}=0.734 \pm 0.054,
\ee
which is consistent with the SM expectation. 
However, this result does not exclude interesting CP violating new physics
effects in other $B$ decays. Since the decay $B \to \psi K_S$  is 
dominated by the tree level $ b \to c\bar c s $ transition in the SM, the
NP contributions to its amplitude are naturally suppressed.
However, at the loop level NP may give large 
contributions to the $B^0- \bar B^0$
mixing as well as to the loop induced decay amplitudes. The former effects are 
universal to all decay modes while the new physics effects to the decay 
amplitudes are nonuniversal and process dependent. Thus the comparison 
of time dependent rate asymmetries in different decay channels, measuring
the same weak phase in the SM, could provide evidence on new physics in the
$B$ meson decay amplitudes. 

$B$ decays involving the $b \to s \bar s s $ transition such as
$B \to \phi K$, $B \to \eta' K$, $\phi K^*$, $\cdots$, proceed through 
the loop induced penguin
diagrams.  These processes provide information on the 
Cabibbo-Kobayashi-Maskawa (CKM) matrix element $V_{ts}$
and are sensitive to physics beyond the SM. They can also be used 
for independent
measurement of the CP violating parameter $\sin(2 \beta)$, and the uncertainty
within the SM, for the decay mode $B \to \phi K_S$ is estimated to be 
\cite{gross1}
\be
|\sin(2 \beta)_{\psi K_S}-\sin(2 \beta)_{\phi K_S}| \lsim {\cal O} (\lambda^2).
\ee
Recently Belle \cite{belle0} and BABAR \cite{babar0} have measured
$\sin(2 \beta)_{\phi K_S}$ with an average
\be
\sin(2 \beta)_{\phi K_S}=-0.39 \pm 0.41,
\ee 
which has a 2.7$\sigma $ deviation 
from the observed value of $\sin(2 \beta)_{\psi K_S}$. The most recent
updated average value of the asymmetry \cite{ave03} is 
\be
\sin(2 \beta)_{\phi K_S}=-0.15 \pm 0.33.
\ee   
Thus the discrepancy between the measured values 
of  $\sin(2 \beta)_{\psi K_S}$ and $\sin(2 \beta)_{\phi K_S}$ may be a 
possible indication of new physics effects in the decay amplitude
of $B$ system.

Recently, several new physics phenomena have been studied to explain the 
above discrepancy \cite{new1,datta1,hiller1,rm1}. If new physics effects 
indeed are present in 
the decay mode $B \to \phi K_S$, then  one can expect to observe similar 
effects in 
other modes having the same internal quark structutre (in fact it is this
speculation which motivated us to undertake the study of this paper).
Therefore, it is also important to explore other signals of new physics 
in order to
corroborate this result. One way to search for new physics effects is
to look for direct CP violation  in decay modes which are having a single
decay amplitude in the SM. It should be reminded here that in order to
observe direct CP violation there should be two interfering decay amplitudes
with different strong and weak phases. An observation of direct CP violation
in such modes,
is an unambiguous signal of new physics. However, non-vanishing of the
direct CP violation requires the relative strong phases between  the SM and NP 
amplitudes to be nonzero. Therefore, if the relative strong phase between 
the two interfering amplitudes is zero, one cannot 
get the new physics information, even if it would be present there. However,
still we have an opportunity which in turn 
can help us to find out the evidence of NP.  
In fact, if one considers $B$ decays to two vector mesons \cite{sinha1} 
then one can show that many signals of new physics effect emerge
including those which are nonzero even if the strong phase difference vanish. 
Therefore, the studies of $B$ meson decaying to two vector 
modes are likely to be the
major contenders to look for NP.

In this paper we intend to study the new physics effects in the decay mode
$\dbbarp \to \phi \kbarp$.~~ Recently, the Belle \cite{belle1} and BABAR
\cite{babar1} collaborations have reported the full angular
analysis of $B \to \phi K^* $ decays. The Belle measurements are as
\bea
&&{ \rm Br} (B^0 \to \phi K^{*0} )=  (10.0 _{-1.5-0.8}^{+1.6+0.7}) 
\times 10^{-6},\nn\\
&&|\hat A_0|^2=0.43 \pm 0.09 \pm 0.04\;,~~~~|\hat \Ape|^2=0.41 \pm 
0.10 \pm 0.04\;,\nn\\
&&\hspace{1.5 cm}
{\rm and} ~~~|\hat \Apa|^2=1-|\hat A_0|^2-|\hat \Apa|^2\;, \nn\\
&&{\rm arg}(\hat \Apa)=-2.57 \pm0.39 \pm 0.09\;, ~~~~{\rm arg}
(\hat \Ape)=0.48 \pm 0.32 
\pm 0.06, \label{pol}
\eea
and the BABAR data are
\bea
&&{\rm Br}(B^0 \to \phi K^{*0} ) =  (11.2 \pm 1.3 \pm 0.8) \times 10^{-6},\nn\\
&&\frac{\Gamma_L}{\Gamma}=0.65 \pm 0.07 \pm 0.02\;,
~~~~A_{CP}=0.04\pm 0.12 \pm 0.02.
\eea
The average branching ratio of Belle and BABAR measurements is given as
\be
{\rm Br}(B^0 \to \phi K^{*0} ) =  (10.7 \pm 1.1) \times 10^{-6}\;.\label{br}
\ee
From the theoretical point of view,
recently, this decay mode has been studied within the SM in the framework of
QCD factorization \cite{cheng01} and PQCD approach \cite{chen02}.
Although the branching ratio in PQCD approach is found to be consistent 
with the experiment, but other observables like the helicity amplitudes
do not agree with the current experimental
data. 
At present,
it appears that  new physics effects indeed 
are present in the $B \to \phi K_S$ mode.
Driven by the experimental activity and the 
possible discrepancy in the CP asymmetry
in the $\phi K_S$ sector,
it is therefore interesting to see the effects of NP in the decay mode
$B \to \phi K^{*0}$, which is our prime objective in this paper.
Here we consider two scenarios beyond the SM, the R-parity violating (RPV)
supersymmetric model and the model with an extra vector-like down quark (VLDQ).
It has been shown very recently that these 
two models can explain the observed $2.7 \sigma$ 
discrepancy in $B \to \phi K_S $ mode \cite{datta1, hiller1, rm1}.

The paper is organised as follows.
Section II includes a general description of the angular distributions and
the observables in $B\to VV$ decays, while in Section \ III we analyze the
particular case of $B^0 \to \phi K^{\ast 0}$ in the SM. The new physics 
effects from VLDQ model and RPV model are considered in
sections III and IV respectively and
in Section\ V we present some concluding remarks.

\section{Observables and angular distributions in $B\to VV$}

Let us consider the decay of a $B$ meson into two vector mesons ($\phi$
and $K^{*}$), followed by the decay $\phi \to K^+ K^-$ and $K^{*0}
\to K^+ \pi ^-$  respectively. Following the
notations of Ref. \cite{chiang1}, the normalized differential angular
distribution can be written as
\begin{eqnarray}
&&\frac{1}{\Gamma} \frac{d^3 \Gamma}{d \cos \theta_1 \l d \cos \theta_{2}
\l d \psi}  =   \frac{9}{8 \pi \Gamma } \;\Bigg\{ L_1 \l \cos^2
\theta_{1} \l \cos^2 \theta_{2} \l + \l \frac{L_2}{2} \l \sin^2 
\theta_{1} \l
\sin^2 \theta_{2} \l \cos^2 \psi \nonumber \\
&& \hspace{2.5 cm}  + \l \frac{L_3}{2} \l
\sin^2 \theta_{1} \l \sin^2 \theta_{2} \l \sin^2 \psi \l + 
\l \frac{L_4}{2
\sqrt{2}} \l \sin 2 \theta_{1} \l \sin 2 \theta_{2} 
\l \cos \psi \nonumber \\
&& \hspace{2.5 cm}  -  \l \frac{L_5}{2 \sqrt{2}} \l \sin 2 \theta_{1} 
\l \sin 2 
\theta_{2}
\l \sin \psi \l - \l \frac{L_6}{2} \l \sin^2 \theta_{1} \l \sin^2 
\theta_{2}
\l \sin 2 \psi \Bigg\}\,,
\label{dg}
\end{eqnarray}
where $\theta_{1}$ ($\theta_2$) is the angle between the 
three-momentum of
$K^+$ ($K^+$) in the $\phi$ ($K^{*0}$) rest frame and the three-momentum of
$\phi$ ($K^{*0}$) in the $B$ rest frame, and in Eq. (\ref{dg}) $\psi$ 
is the angle between 
the normals to the planes defined by 
 $K^+ K^-$ and $K^+ \pi^-$, in the $B$
rest frame. The coefficients $L_i$ can be expressed in terms of three
independent amplitudes, $A_0$, $A_\|$ and $A_\bot$, which correspond to
the different polarization states of the vector mesons $\phi$ and
$K^{*0}$ as
\begin{eqnarray}
L_1 = |A_0|^2\;, \hspace{2.cm} & L_4 = {\rm Re} [A_{\|} A^*_0]\;, \nonumber
\\ & & \nonumber \\ L_2 = {|A_{\|}|}^2\;, \hspace{2.cm} & L_5 = {\rm Im}
[A_\bot A^*_0]\;, \nonumber \\ & & \nonumber \\ L_3 = |A_{\bot}|^2\;,
\hspace{2.cm} & L_6 = {\rm Im} [A_{\bot} A^*_{\|}]\;. \label{K}
\end{eqnarray}
In the above $A_0$, $\Apa$, and $\Ape$ are complex amplitudes of the three
helicity states in the transversity basis.
 The CP odd and CP even 
fractions of the decay 
$B \to \phi K^{*0}$ are given by $|\Ape|^2$ and $(|A_0|^2+|\Apa|^2 )$
respectively.

It should be noted here that only six of the nine
possible observables given by the squared amplitude $A^\ast A$ can be
measured independently. This is because of the fact that both the 
daughter vector mesons ($\phi$ and $K^*$)
are considered to decay into two spin zero particles.

The decay mode $B\to V_1V_2$ can also be described
in the helicity basis, where 
the amplitude for the helicity
matrix element can be parametrized 
as~\cite{Kra92}
\begin{eqnarray}
H_{\lambda}&=& \langle V_1 (\lambda)V_2(\lambda)|{\cal H}_{eff} |B^0 
\rangle\nonumber\\
&=& \varepsilon_{1 \mu}^* (\lambda) \l \varepsilon_{2
\nu}^* (\lambda) \left [ a g^{\mu \nu} + \frac{b}{m_1 m_2} p^{\mu}
p^{\nu} + \frac{i c}{m_1 m_2} \epsilon^{\mu \nu \alpha \beta} p_{1
\alpha} p_{\beta} \right ]\;,
\label{hlam}
\end{eqnarray}
where $p$ is the $B$ meson momentum and $\lambda =0, \pm 1$ are 
the helicity of both the
vector mesons. In the above expression $m_i$, $p_i$ and $\varepsilon_i$ 
($i=1,2$) stand for their
masses, momenta and polarization vectors respectively. 
Furthermore, the three invariant amplitudes $a$, $b$, and $c$ are related to
the helicity amplitudes by
\begin{equation}
H_{\pm 1} = a \pm c \l \sqrt{x^2 - 1}\;,
\esp
H_0 = - a x - b \l (x^2 - 1)\;,
\label{a}
\end{equation}
where $x =(p_1 \cdot p_2)/m_1 m_2 = (m_B^2 - m^2_1 - m^2_2)/(2 m_1 m_2)$. 

The corresponding decay rate using the helicty basis amplitudes can be given as
\begin{equation}
\Gamma = \frac{p_{cm}}{8 \pi m_B^2} \biggr( |H_0|^2+|H_{+1}|^2 +|H_{-1}|^2 
\biggr)\;,
\end{equation}
where $p_{cm}$ is the magnitude c.o.m. momentum of the outgoing vector 
particles. 
It is also conveninet to express the relative decay rates into $V$ meson 
states with longitudinal and transverse
polarizations as
\begin{eqnarray}
\frac{\Gamma_L}{\Gamma_0} &  = &
\frac{|H_0|^2}{|H_0|^2+|H_{+1}|^2+|H_{-1}|^2}
\;\; , \nonumber \\
\frac{\Gamma_T}{\Gamma_0} &  = &
\frac{|H_{+1}|^2+|H_{-1}|^2}{|H_0|^2+|H_{+1}|^2+|H_{-1}|^2}
\;\; .
\end{eqnarray}

The  amplitudes in  transversity  and helicity basis are related to each other
through the following relations 
\begin{eqnarray}
A_{\bot} \l = \l \frac{H_{+1} - H_{-1}}{\sqrt{2}}, \esp
A_{\|} \l = \l \frac{H_{+1} + H_{-1}}{\sqrt{2}}, \esp A_0 \l =
\l H_0 \label{cb}.
\end{eqnarray}

Correspondingly, the coefficients $L_i$ can also be written in terms 
of the parameters $a$, $b$, and $c$ as
\begin{eqnarray}
\lefteqn{L_1 = |x a + (x^2-1) b|^2} \hspace*{5.5cm} & &
L_4 = -\sqrt{2}\,\left[ x\, |a|^2 + (x^2-1)\,{\rm Re}(a b^*)\right],
\nonumber \\
\lefteqn{L_2 = 2\,|a|^2} \hspace*{5.5cm} & &
L_5 = -\sqrt{2\,(x^2-1)} \left[ x\, {\rm Im} (a^\ast c)
+ (x^2-1)\,{\rm Im}(b^\ast c)\right], \nonumber \\
\lefteqn{L_3 = 2\,(x^2-1)|c|^2} \hspace*{5.5cm} & &
L_6 = 2\sqrt{x^2-1}\; {\rm Im} ( a^\ast c).
\end{eqnarray}

Similar to the $H_{\lambda}$ amplitudes, one can also
write the corresponding amplitudes for the complex conjugate process.
The helicity amplitudes $\bar H_\lambda$ for the decay $\bar B \to \bar 
V_1 \bar V_2$, where $\bar V_1$ and $\bar V_2$ are the antiparticles of
$V_1$ and $V_2$ respectively, have the same decomposition with
\begin{equation}
a \to \bar a, ~~~~b \to \bar b ~~~~{\rm and}~~~c \to -\bar c.
\end{equation}
Here in general, the parameters $a$, $b$, and $c$ are complex numbers. 
One can then write these amplitudes as 
\begin{equation}
a = |a| \; e^{i \, (\delta_a + \varphi_a)}\;,
\label{contrib}
\end{equation}
where $\delta$ and $\varphi$ stand for ``strong'' (CP-conserving) and
``weak'' (CP-violating) phases respectively.
In fact, the parameter $a$ can have contributions from different interfering 
decay amplitudes.
Since the decay $ B \to \phi K^{*0}$ receives dominant contribution 
only from the one loop $ b \to s \bar s s$ penguin diagram with 
top quark in the loop (i.e, it is described
by a single weak decay amplitude), we consider only one term in the 
amplitude $a$. 
The $ \bar a $ then can be obtained from
$a$, by changing the sign of weak phase. 
Similar relations as that of (\ref{contrib})
can be written for parameters $b$ and $c$.

In our analysis, we will take into account both the decay $B^0\to \phi
K^{\ast 0}$ and its CP-conjugate process, $ \bar B^0\to\phi \bar K^{\ast 0}$.
In fact, there are several ways for CP violation to manifest itself. 
But, the most familiar 
one is in the partial rate asymmetries. Since there are three differential 
decay amplitudes, the partial rate asymmetries may show up in either of them.
These asymmetries can be studied by measuring the coefficients
of first three terms in Eq.(\ref{dg})  for $B^0$ and $\bar B^0$ decays
and comparing those coefficients. 
In addition to these, CP violation can also be observed 
in the interfering amplitudes, i.e., in the measurement of 
the coefficients of last three terms of
Eq. (\ref{dg}). Notice, however, that without separating the observables of 
$B^0$ and $\bar B^0$ 
decays, the relevant information
on CP violation cannot be extracted. Therefore, one must obtain the angular
distribution for $B^0 \to \phi K^{* 0}$ and $\bar B^0 \to \phi 
\bar K^{*0}$ decays separetely and determine the coefficients $L_{1-6}$
in each case.

Now, in principle, from the angular analysis of $\dbbarp \to\phi
\kbarp$~~ decays one can measure twelve observables, these are in fact the 
coefficienets $L_i$ and $\bar
L_i$ with $i=1$ to 6. The CP violating effects in these observables are
given as
\bea
C_1 = L_1 -\bar L_1 &= &-4x(x^2-1) |a|~|b| \sin \dab \sin \pab, \nn\\
C_2 = L_2 -\bar L_2 &= & 0, \nn\\
C_3 = L_3 -\bar L_3 &= & 0, \nn\\
C_4 = L_4 -\bar L_4 &= &2 \sqrt{2}(x^2-1) |a|~|b| \sin \dab \sin \pab, \nn\\
C_5 = L_5 + \bar L_5 &= &-2 \sqrt{2(x^2-1)} |c|\biggr[
x|a| \cos \dca \sin \pca +
(x^2-1) |b| \cos \dcb \sin \pcb \biggr],\nn\\
C_6 = L_6 +\bar L_6 &= &4 \sqrt{(x^2-1)} |a|~|c| \cos \dca \sin \pca, 
\label{cpv1}
\eea
where $\delta_{ij}=\delta_i-\delta_j$ and $\varphi_{ij}=\varphi_i-\varphi_j$.
It is important to note that the CP violating observables $C_5$ and $C_6$
do not require FSI strong phase differences and especially sensitive to CP
violating weak phases. 

So far, we have limited our discussion to that of 
within the framework of the SM and 
presented various combinations 
in which CP violation effects will show up.
We are now ready to explore the effects of new physics. 
 
Now in the presence of new physics the total invariant amplitude 
may be written as
\be
a_T=a_{SM}+a_{NP}=a_{SM}\big[1+r_a e^{i(\da^n +\pa^n)}\big],
\ee
where $r_a=|a_{NP}/a_{SM}|$, ($a_{SM}$ and $a_{NP}$ correspond to the
SM and NP amplitudes) and $\da^n$ ($\pa^n$) is the relative strong (weak)
phase between the SM and NP amplitudes. Similar expressions can be written for
the other two amplitudes $b$ and $c$. 

Incorporating the generic new physics contribution we write the modified 
observables and after some algebra we arrive at the new $C$'s as given below.
Thus, in the presence of new physics the CP violating observables $C_{(1-6)}$
read as
\bea
C_1 &= & -4 \biggr[x (x^2-1)|a|~|b| \sin \dab \sin \pab
+x^2 |a|^2r_a \sin
\da^n \sin\pa^n +(x^2-1)^2 |b|^2 r_b \sin
\db^n \sin\pb^n\nn\\
&& \hspace{0.3 cm}+
x(x^2-1)|a|~|b| \Big( r_a \sin
(\dab+\da^n) \sin (\pab +\pa^n)
+ r_b \sin(\dab+
\db^n )\sin (\pab+ \pb^n)\nn\\
&&\hspace{0.3 cm}
+ r_a r_b \sin(\dab+
\dab^n) \sin (\pab+\pab^n) \Big) \biggr],\nn\\
C_2 &=&  -8 |a|^2 r_a \sin \da^n \sin \pa^n, \nn\\
C_3 &=& -8(x^2-1)|c|^2 r_c \sin
\dc^n \sin\pc^n, \nn\\
C_4 &= & 2 \sqrt 2 \biggr[(x^2-1) |a|~|b| \Big( \sin \dab \sin \pab
+r_a \sin(\dab+\da^n) \sin (\pab +\pa^n)\nn\\
&&\hspace{0.3 cm}+ r_b \sin(\dab+
\db^n )\sin (\pab+ \pb^n)
+r_a r_b \sin(\dab+
\dab^n) \sin (\pab+\pab^n) \Big)\nn\\
&&\hspace{0.3 cm}+
2|a|^2x r_a \sin
\da^n \sin\pa^n
 \biggr],
\nn\\
C_5 &=& -2 \sqrt{2(x^2-1)} |c|\biggr[ x |a| 
\Big(\cos\dca \sin \pca+r_c \cos (\dca+\dc^n) \sin 
(\pca+\pc^n)\nn\\
&& \hspace{0.3 cm}+r_a \cos (\dca-\da^n) \sin 
(\pca-\pa^n)+
r_a r_c \cos
(\dca+\dca^n) \sin(\pca+\pca^n)\Big)\nn\\
&& \hspace{0.3 cm}+ (x^2-1)|b|
\Big(\cos\dcb \sin \pcb+r_c \cos (\dcb+\dc^n) \sin 
(\pcb+\pc^n)\nn\\
&& \hspace{0.3 cm}+r_b \cos (\dcb-\db^n) \sin 
(\pcb-\pb^n)
+r_b r_c \cos
(\dcb+\dcb^n) \sin(\pcb+\pcb^n)\Big)
\biggr],\nn\\
C_6 &=& 4 \sqrt{(x^2-1)}|a|~ |c| \biggr[  
\cos\dca \sin \pca+r_c \cos (\dca+\dc^n) \sin 
(\pca+\pc^n)\nn\\
&& \hspace{0.3 cm}+r_a \cos (\dca-\da^n) \sin 
(\pca-\pa^n)+
r_a r_c \cos
(\dca+\dca^n) \sin(\pca+\pca^n)
 \biggr]\;,\label{cpv2}
\eea
where $\delta_{ij}^n=\delta_i^n-\delta_j^n$ and $\varphi_{ij}^n
=\varphi_i^n-\varphi_j^n$.
After obtaining the expressions for the CP violating observables, in the 
presence of new physics, we now proceed to explore the specific cases.
Before going to do that let us now first look at the
relevant quantities in the SM. The discrepancy, if any, in the SM will
necessitate the inclusion of NP to explain the same.   

\section{ SM contribution to the amplitude
$B^0\to\phi K^{\ast 0}$}

Let us now focus on the decay $\bar B^0 \to\phi \bar K^{\ast 0}$. 
In the SM, this decay process proceeds through the
quark level transition $b \to s \bar s s$, which is induced by the
QCD, electroweak (EW) and magnetic penguins. QCD penguins with the top
quark in the loop contribute predominantly to such process.
However, since we are looking for NP, here we would like to retain
all the contributions. The effective Hamiltonian describing the
decay $b\to s\bar ss$ \cite{flei1} is given as
\be 
H_{eff}=
-\frac{G_F}{\sqrt{2}}V_{tb}V_{ts}^* \left( \sum_{j=3}^{10}C_j O_j
+C_g O_g \right),
\ee
where $O_3, \cdots, O_{6}$ and $O_7,
\cdots, O_{10}$ are the standard QCD and EW penguin operators,
respectively, and $O_g$ is the gluonic magnetic operator. Within
the  SM and at scale $M_W$, the Wilson coefficients $C_1(M_W),
\cdots , C_{10}(M_W)$ at next to leading logarithmic order (NLO)
and $C_g(M_W) $ at leading logarithmic order (LO) have been given
in Ref. \cite{Buc96}. The corresponding QCD corrected values at the energy
scale $\mu=m_b$ can be obtained using the renormalization group
equation, as described in Ref. \cite{Ali98}.

To calculate the $B$ meson decay rate, we use the factorization
approximation to evaluate the hadronic matrix element $\langle O_i
\rangle\equiv\langle \bar K^{*0}\phi|O_i|\bar B^0 \rangle$. 

For evaluating the matrix element of
the most relevant operator, i.e., $O_g$, we use the procedure of
\cite{bar1}, where it has been shown that the operator $O_g$ is
related to the matrix element of the QCD and electroweak penguin
operators as
\bea
 \langle O_g\rangle &=&-\frac{\alpha_s}{4\pi}\frac{m_b}{\sqrt{
 \langle q^2 \rangle }}\left [
 \langle O_4\rangle + \langle O_6\rangle
 -\frac{1}{N_C}(\langle O_3\rangle+\langle O_5\rangle)\right ].
\eea
In the above equation $q^\mu$ is the momentum transferred by the gluon to the
$(\bar s,s)$ pair. The parameter $\langle q^2\rangle$ introduces
certain uncertainty into the calculation. In the literature its
value is taken in the range $1/4 \lsim \langle q^2 \rangle/m_b^2
\lsim 1/2$ \cite{des1}, and we will use $\langle q^2 \rangle/m_b^2
= 1/2$ \cite{Ali98} in our numerical calculations.

Thus, in the factorization approach, the amplitude $A\equiv\langle
\phi \bar K^{*0} |H_{eff}|\bar B^0 \rangle$ of the decay  
$\bar B^0\to \phi \bar K^{*0}$ takes a form
\bea
A(\bar B^0 \to \phi \bar K^{*0})=-\frac{G_F}{\sqrt{2}}V_{tb}V_{ts}^* \left[
a_3+a_4+a_5-\frac{1}{2}(a_7+a_9+a_{10}) \right] X\;,
 \label{A}
 \eea
where 
\be
X \equiv \langle \phi(\varepsilon_2, p_2)|\bar s \gamma_\mu (1-\gamma_5)
s | 0 \rangle\langle \bar K^{*0}(\varepsilon_1, p_1)|\bar s \gamma^\mu 
(1-\gamma_5)
b | \bar B^0(p) \rangle \label{fc}\ee 
stands for the factorizable hadronic matrix element.
The coefficients $a_i$ are given by
\bea a_{2i-1}=
C^{eff}_{2i-1}+\frac{1}{N_c}C^{eff}_{2i}\,, \qquad\qquad a_{2i}=
C^{eff}_{2i}+\frac{1}{N_c}C^{eff}_{2i-1}\,,
\eea
where $N_C$ is the number of colors. 

In the factorization approximation the factorized matrix element 
$X$ (Eq. (\ref{fc})) can be written, in general, in terms of form factors 
and decay constants. These are defined as \cite{bsw}
\begin{eqnarray}
\langle \phi (\varepsilon_2,p_2) | V_\mu | 0 \rangle & = & f_\phi \, m_\phi \,
\varepsilon^\ast_{2 \mu}, \nonumber \\
\langle \bar K^{* 0} (\varepsilon_1, p_1) | V_\mu | \bar B (p) \rangle & = 
& -\,\frac{2}{m_{K^*} +
m_B} \; \epsilon_{\mu\nu\alpha\beta} \, \varepsilon_1^{\ast\,\nu} \,
p^\alpha {p_1}^\beta \, V(q^2), \nonumber \\
\langle \bar K^{*0} (\varepsilon_1, p_1) | A_\mu | \bar B (p) \rangle & = & i\,
\frac{2\, m_{K^*} (\varepsilon_1^\ast\cdot q)}{q^2}\;
q_\mu\; A_0(q^2) +
i\, (m_{K^*} + m_B) \left[\varepsilon_{1 \mu}^\ast - \frac{(\varepsilon_1^\ast
\cdot q)}{q^2}\; q_\mu \right] A_1 (q^2) \nonumber \\
& & -\, i\, \left[(p + p_1)_\mu\, - \frac{(m_B^2-m_{K^*}^2)}{q^2}
\;q_\mu \right] \frac{(\varepsilon_1^\ast \cdot q)}{m_{K^*} + m_B} \;
A_2(q^2)\;,
\end{eqnarray}
where $V_\mu$ and $A_\mu$ are the corresponding vector and
axial-vector quark currents and $q=p-p_1$ is the momentum transfer. The
vector and axial-vector form factors can be estimated from the analysis
of semileptonic $B$ decays, using the ansatz of pole dominance to
account for the momentum dependences in the region of interest.

In this way the invariant amplitudes $a$, $b$, and $c$ read as
\begin{eqnarray}
a & = & i ~P_{eff}\,f_\phi~ m_\phi \, ( m_B + m_{K^\ast} )\, 
A_1^{B\to K^\ast}(m_\phi^2),
\nonumber \\
b & = & -i~ P_{eff}\,f_\phi~ m_\phi \, \left(\frac{2\, m_{K^\ast}\, m_\phi}
{m_B + m_{K^\ast}} \right)  A_2^{B\to K^\ast}(m_\phi^2), \nonumber \\
c & = & -i~ P_{eff}\, f_\phi ~m_\phi \, \left(\frac{2\, m_{K^\ast}\, m_\phi}
{m_B + m_{K^\ast}} \right)  \, V^{B\to K^\ast}(m_\phi^2),
\label{peng}
\end{eqnarray}
where
\begin{equation}
P_{eff} = \frac{G_F}{\sqrt{2}} \; V^\ast_{ts}\, V_{tb} \left[a_3 + a_4
+ a_5 - \frac{1}{2} ( a_7 + a_9 + a_{10} )\right]\,.
\label{ceffp}
\end{equation}

The values of the QCD improved effective
coefficients $a_i$ can be found in Ref. \cite{Ali98}. Now
substituting the values of $a_i$ for $N_C$=3, from Ref. \cite{Ali98},
the value of the form factor $V^{B \to K^*}(m_\phi^2)=$ 0.38, 
$A_1^{B \to K^*}(m_\phi^2)=A_2^{B \to K^*}(m_\phi^2)=$ 0.34,
and using
the $\phi$ meson decay constant $f_{\phi}=$ 0.233 GeV and $\tau_{B^0}=
1.542 \times 10^{-12}$ sec \cite{pdg}, we obtain
the branching ratio in the SM as
\be
BR^{SM}(\bar B \to \phi \bar K^{*0})=8.32
\times 10^{-6},
\ee
which is slightly below the experimental
limit (Eq. (\ref{br})).

However, the normalized polarization amplitudes (i.e, $|\hat A_i|^2
=\frac{|A_i|^2}{(|A_0|^2+|\Apa|^2+|\Ape|^2)}$ with $i=0,\parallel,\perp$) 
obtained are
\be
|\hat A_0|^2=0.869\;,~~~|\hat \Ape|^2=0.048\;,~~~~|\hat \Apa|^2=0.083\;,
\ee
which do not agree with the present experimental data (Eq. (\ref{pol})). 
Furthermore, within 
the SM all the three invariant amplitudes $a$, $b$, and $c$ have the 
vanishing weak phase (i.e., phase of $V_{tb}V_{ts}^*$).
In the
framework of the factorization approximation strong phases are
originated by the final state interactions
\cite{bss}. Moreover, 
these phases are the same for the amplitudes
$a$, $b$, and $c$, since the combination of $a_i$ coefficients in all
cases is the same, as is encoded in $P_{eff}$. Also we have the same
$P_{eff}$ appearing in all the amplitudes.
In this way, within the factorization
approximation we have
\begin{equation}
\delta_a=\delta_b=\delta_c = \arg \left[a_3 + a_4
+ a_5 - \frac{1}{2} ( a_7 + a_9 + a_{10} )\right]\equiv\delta\,.
\end{equation}
Thus in the SM, the amplitudes $a$, $b$ and $c$ have the same
strong and vanishing weak phases. Therefore, all the CP violating parameters 
$C_{1-6}$ in Eq. (\ref{cpv1}) are identically 
zero in the SM.
So a nonzero observation of CP violation, in these observables, is a 
clear signal of new physics.

\section{Contribution from the VLDQ model}

Now we consider the model with an additional vector-like down
quark \cite{ref11}. It is a simple model beyond the SM with an
enlarged matter sector with an additional vectorlike down quark
$D_4$. The most interesting effects in this model concern CP
asymmetries in neutral $B$ decays into final CP eigenstates. At a
more phenomenological level, models with isosinglet quarks provide
the simplest self-consistent framework to study deviations of $3
\times 3$ unitarity of the CKM matrix as well as allow flavor
changing neutral currents at the tree level. The presence of an
additional down quark implies a $4 \times 4$ matrix $V_{i \alpha}$
$(i=u,c,t,4,~\alpha= d,s,b,b')$, diagonalizing the down quark mass
matrix. For our purpose, the relevant information for the low
energy physics is encoded in the extended mixing matrix. The
charged currents are unchanged except that the $V_{CKM}$ is now the $3
\times 4$  upper submatrix of $V$. However, the distinctive feature
of this model is that the FCNC enters neutral current Lagrangian of
the left handed down-quarks:
 \be
 {\cal L}_Z= \frac{g}{2 \cos
\theta_W} \left [ \bar u_{Li} \gamma^{\mu} u_{Li} - \bar d_{L
\alpha}U_{\alpha \beta} \gamma^\mu d_{L \beta}-2 \sin^2 \theta_W
J_{em}^\mu \right ]Z_{\mu},
\ee
with
\be
U_{\alpha \beta}
=\sum_{i=u,c,t} V_{\alpha i}^\dagger V_{i \beta} =\delta_{\alpha
\beta} - V_{4 \alpha}^* V_{4 \beta},
\ee
where $U$ is the neutral
current mixing matrix for the down sector, which is given above. As
$V$ is not unitary, $U \neq {\bf{1}}$. In particular its
nondiagonal elements do not vanish:
\be
U_{\alpha \beta}= -V_{4
\alpha}^* V_{4 \beta} \neq 0~~~{\rm for}~ \alpha \neq \beta.
\ee

Since the various $U_{\alpha \beta}$ are nonvanishing, they would
signal new physics and the presence of the FCNC at the tree level, and this
can substantially modify the predictions for CP asymmetries. The
new element $U_{s b}$ which is relevant to our study is given as
\be
U_{sb}= V_{us}^* V_{ub}+V_{cs}^*V_{cb}+V_{ts}^*V_{tb}.
\ee

The decay mode $B^0 \to \phi K^*$ receives new contributions both from 
the color allowed as well as the color 
suppressed $Z$-mediated FCNC transitions.
The new additional operators are given as
\bea
&&O_1^{Z-FCNC}=[\bar s_\alpha \gamma^\mu (1-\gamma_5) b_\alpha][\bar s_\beta
\gamma_\mu(C_V^s -C_A^s \gamma_5) s_\beta],\nn\\
&&O_2^{Z-FCNC}=[\bar s_\beta \gamma^\mu (1-\gamma_5) b_\alpha][\bar s_\alpha
\gamma_\mu(C_V^s -C_A^s \gamma_5) s_\beta],
\eea
where $C_V^s$ and $C_A^s$ are the vector and axial vector $Z s \bar s$
couplings. Using the Fierz transformation and the identity
$(C_V^s-C_A^s \gamma_5)=[(C_V^s+C_A^s)(1- \gamma_5)+(C_V^s-C_A^s)(1+ \gamma_5)]
/2$ as done in Ref. \cite{rm1} for $B \to \phi K_S$ mode, 
the invariant amplitudes $a$, $b$, and $c$ are 
given in the factorization approximation as
\begin{eqnarray}
a_{NP} & = & i ~\frac{G_F}{\sqrt 2} U_{sb}~2 (C_V^s +\frac{C_A^s}{3})\,
f_\phi~ m_\phi \, 
( m_B + m_{K^\ast} )\, 
A_1^{B\to K^\ast}(m_\phi^2),
\nonumber \\
b_{NP} & = & -i~ \frac{G_F}{\sqrt 2} U_{sb}~2 (C_V^s +\frac{C_A^s}{3})
\,f_\phi~ m_\phi \, \left(\frac{2\, m_{K^\ast}\, m_\phi}
{m_B + m_{K^\ast}} \right)  A_2^{B\to K^\ast}(m_\phi^2), \nonumber \\
c_{NP} & = & -i~ \frac{G_F}{\sqrt 2} U_{sb}~2 (C_V^s +\frac{C_A^s}{3})\, f_\phi ~m_\phi \, \left(\frac{2\, m_{K^\ast}\, m_\phi}
{m_B + m_{K^\ast}} \right)  \, V^{B\to K^\ast}(m_\phi^2).
\label{peng1}
\end{eqnarray}

The values for $C_V^s$ and $C_A^s$ are taken as
\be
C_V^s= -\frac{1}{2}+\frac{2}{3} \sin^2 \theta_W\;,
~~~~~~~~~C_A^s=-\frac{1}{2}\;.
\ee
Now using  $\sin^2 \theta_W$=0.23 and
 the value of $U_{bs}$ as
\be
|U_{bs}| \simeq 1 \times 10^{-3}\;,
\ee 
 we find 
\be 
r \equiv r_{a, b, c}\simeq 0.6.
\ee
The relative strong and weak phases in this model turn out to be same
as all the amplitudes receive the same contribution from new physics effects.
Thus we consider  $\da^n=\db^n=\dc^n= \delta^n$ and
$\pa^n=\pb^n=\pc^n=\varphi^n$ in our analysis. The branching ratio  
turns out to be
\be
{\rm Br}(\bar B^0 \to \phi \bar K^{*0})={\rm Br}^{SM}(1+r^2+2r \cos\phi_{NP}),
\ee
where $\phi_{NP} = (\delta^n+\varphi^n) $ and ${\rm Br}^{SM}$
denotes the branching ratio in the SM. Now if we plot (Fig-1) 
the branching ratio
vs. $\phi_{NP}$, we see that the observed branching ratio can be easily 
accomodated in this model. However, it should be noted that, since new
physics contributions to all the 
three amplitudes $a$, $b$ and $c$ are identical, the values of the
normalized polarization amplitudes will remain same as their SM values. 


\begin{figure}[htb]
   \centerline{\epsfysize 2.5 truein \epsfbox{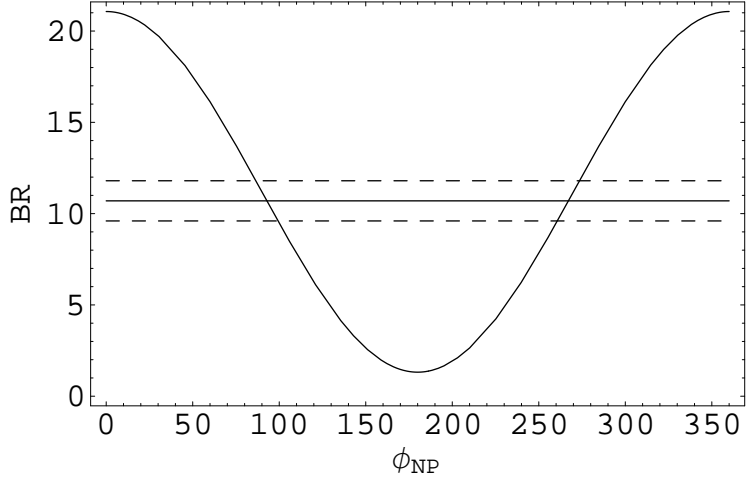}}
   \caption{
 Branching ratio of $B \to \phi K^{*0}$ process 
(in units of $10^{-6}$) versus the phase
$\phi_{NP}$ (in degree).
The horizontal solid line is the central experimental
value whereas the dashed horizontal lines denote the error
limits.}
 \end{figure}

The CP violating observables are obtained from Eq. (\ref{cpv2}) as
\bea
C_1 &=& =-4 \left [ x |a|+(x^2-1) |b| \right ]^2 
r \sin \delta^n \sin \varphi^n, \nn\\
C_2 &=& -8 |a|^2 r \sin \delta^n \sin \varphi^n, \nn\\
C_3 &=& -8 (x^2-1) |c|^2 r \sin \delta^n \sin \varphi^n, \nn\\
C_4 &=&4 \sqrt 2 \left [  |a|^2 x +(x^2-1) |a|~|b| \right ]
r \sin \delta^n \sin \varphi^n, \nn\\
C_5 &= & 0,\nn\\
C_6 &= & 0. \label{eq:cp1}
\eea
From the above expressions it can be noted that
new physics effects from VLDQ model predict nonzero values for 
the CP violating parameters $C_{(1-4)}$.
Thus, observation of the non-vanishing CP asymmetries implies, in general,
the new physics interaction  should be of $(V-A)(V-A)$ form, as in the
VLDQ model. However, in order to be different from zero, the CP violating
parameters $C_{(1-4)}$ require the presence of nonzero relative strong phases
$\delta^n$. So in case, these relative strong phases turn out to be zero 
or too small then the  asymmetries in (\ref{eq:cp1}) could be too small to
be observed experimentally, even in the presence of new physics.

\section{Contribution from the R-Parity violating supersymmetric model}

We now analyze the decay mode in minimal supersymmetric model with
R-parity violation \cite{gue1}.
In the supersymmetric models there may be interactions which
violate the baryon number $B$ and the lepton number $L$
generically. The simultaneous presence of both $L$ and $B$ number
violating operators induce rapid proton decay, which may contradict
strict experimental bound. In order to keep the proton lifetime
within experimental limit, one needs to impose additional symmetry
beyond those of the SM gauge symmetry. This is to force the 
unwanted baryon and lepton
number violating interactions to vanish. In most cases this has
been done by imposing a discrete symmetry called R-parity, defined
as, $R_p=(-1)^{(3B+L+2S)}$, where $S$ is the intrinsic spin of the
particles. Thus the $R$-parity can be used to distinguish the
particle ($R_p$=+1) from its superpartner ($R_p=-1$). This
symmetry not only forbids rapid proton decay but also renders
the stablility of the lightest supersymmetric particle (LSP). However, this
symmetry is ad hoc in nature. There is no theoretical arguments in
support of this discrete symmetry. Hence, it is interesting to see
the phenomenological consequences of the breaking of R-parity in
such a way that either $B$ and $L$ number is violated, both are
not simultaneously violated, thus avoiding rapid proton decays.
Extensive studies have been done to look for the direct as well as
indirect evidence of R-parity violation from different processes
and to put constraints on various R-parity violating couplings.
The most general $R$-parity and Lepton number violating
super-potential is given as
\begin{equation}
W_{\not\!{L}} =\frac{1}{2} \lambda_{ijk} L_i L_j E_k^c
+\lambda_{ijk}^\prime L_i Q_j D_k^c \;,\label{eq:eqn10}
\end{equation}
where, $i, j, k$ are generation indices, $L_i$ and $Q_j$ are
$SU(2)$ doublet lepton and quark superfields and $E_k^c$, $D_k^c$
are lepton and down type quark singlet superfields. Further,
$\lambda_{ijk}$ is antisymmetric under the interchange of the
first two generation indices. Thus the relevant four fermion
interaction induced by the R-parity and lepton number violating
model is \cite{gue1}
\bea
{\cal H}_{\not\!{R}}  = 
\frac{1}{8 N_c m^2_{\tilde \nu}}\biggr[
&(\lambda_{i23}^{\prime *} \lambda_{i22}^{\prime})&
(\bar s \gamma^\mu(1-\gamma_5)s)(\bar s \gamma_\mu (1+\gamma_5)b)
\nn\\
+&
(\lambda_{i32}^{\prime} \lambda_{i22}^{\prime *})&
(\bar s \gamma^\mu(1+\gamma_5)s)(\bar s \gamma_\mu (1-\gamma_5)b)
\biggr].\label{Ha}
\eea
Using factorization approximation, the amplitudes
$a$, $b$ and $c$ in R-parity violating model are obtained as
\bea
a_{NP} &=& i f_\phi m_{\phi} (m_B + m_{K^*})A_1(m_\phi^2)
 \biggr[\frac{1}{8 N_c m^2_{\tilde \nu}}
\Big((\lambda_{i32}^{\prime } \lambda_{i22}^{\prime *})
-(\lambda_{i23}^{\prime *} \lambda_{i22}^{\prime})\Big)\biggr],\nn\\
b_{NP} &=&- i f_\phi m_{\phi} \left (\frac{2m_\phi m_{K^*}}{m_B + m_{K^*}}
\right )
A_2(m_\phi^2)\biggr[
 \frac{1}{8 N_c m^2_{\tilde \nu}}
\Big((\lambda_{i32}^{\prime } \lambda_{i22}^{\prime *})
-(\lambda_{i23}^{\prime *} \lambda_{i22}^{\prime})\Big)\biggr],\nn\\
c_{NP} &=&- i f_\phi m_{\phi} \left (\frac{2 m_\phi m_{K^*}}{m_B + m_{K^*}}
\right )V(m_\phi^2)
\biggr[\frac{1}{8 N_c m^2_{\tilde \nu}}
\Big((\lambda_{i32}^{\prime } \lambda_{i22}^{\prime *})
+(\lambda_{i23}^{\prime *} \lambda_{i22}^{\prime})\Big)\biggr],
\eea
where the summation over $i=1,2,3$ is implied. Notice, however, that 
in this case
the contributions from NP to the amplitudes $a$ and $b$ are same
while the contribution to amplitude $c$ is different.
Now considering $r \equiv r_a =r_b $, $\delta^n \equiv \da^n=\db^n$
and $\varphi^n \equiv \pa^n=\pb^n$, the CP violating observables 
as obtained from (\ref{cpv2}) are given as
 
\bea
C_1 &=& -4 \left [ x |a|+(x^2-1) |b| \right ]^2 
r \sin \delta^n \sin \varphi^n, \nn\\
C_2 &=& -8 |a|^2 r \sin \delta^n \sin \varphi^n, \nn\\
C_3 &=& -8 (x^2-1) |c|^2 r \sin \delta_c^n \sin \varphi_c^n, \nn\\
C_4 &=&4 \sqrt 2 \left [  |a|^2 x +(x^2-1) |a|~|b| \right ]
r \sin \delta^n \sin \varphi^n, \nn\\
C_5 &= & -2\sqrt{2(x^2-1)}\Big[ x |a|~|c| +(x^2-1) |b|~|c| \Big]\nn\\
&& \hspace{0.2 cm} \times
\Big[r r_c \cos \dca^n \sin \pca^n +r_c \cos \dc^n \sin \pc^n
-r \cos \delta^n \sin \varphi^n \Big],\nn\\
C_6 &= & 4 \sqrt{(x^2-1)} |a|~|c|
\Big[r r_c \cos \dca^n \sin \pca^n +r_c \cos \dc^n \sin \pc^n
-r \cos \delta^n \sin \varphi^n \Big].
 \label{eq:cp2}
\eea
This set of equations deserves some attention.It should be noted here that
the observables $C_{5,6}$ come with cosine of the relative strong phase. Thus,
the nonvanishing of $C_{5,6}$ (even in the vanishing relative strong phase limit) imply the presence of new physics effects from
R-parity violating model or models with $(V-A)(V+A)$ interaction Hamiltonian.
Furthermore, in this case one can get the new physics signal even 
with vanishing relative strong phases between the SM and NP contributions.

To have an idea about the maginitude of new physics contributions arising 
from R-parity violating model, we consider the values of R-parity couplings 
from \cite{bdutta}
as
\be
\frac{1}{8 m^2_{\tilde \nu}}
\Big(\lambda_{i32}^{\prime } \lambda_{i22}^{\prime *}\Big)= ke^{-i \theta}
~~~~{\rm and} ~~~~\frac{1}{8 m^2_{\tilde \nu}}
\Big(\lambda_{i23}^{\prime *} \lambda_{i22}^{\prime}\Big)= -ke^{-i \theta},
\ee
where $k$ is the magnitude and $\theta $ is the new weak phase of
R-parity violating couplings. For $|\lambda_{322}^\prime|=
|\lambda_{332}^\prime|=|\lambda_{323}^\prime|=0.055$, $\tan \theta$=0.52 and
sneutrino mass $m_{\tilde \nu}=200$ GeV, we obtain for $N_C=3 $,
\be
r\equiv r_a=r_b=0.43 ~~~{\rm and}~~~r_c=0.16.
\ee

\section{Conclusions}

We study the decay process $B^0 \to \phi K^{*0}$, showing that
the analysis of the final outgoing particles can be used to detect
the presence of new physics. If there happens to be a new physics
contribution to its decay amplitude, with a different weak phase, 
then the
standard technique for detecting such NP effects is by measuring
direct CP asymmetry parameters. However, the nonvanishing value of
these parameters require nonzero relative strong phase
between SM and NP amplitudes. So if the strong phases
of the SM and NP amplitudes turn out to be equal, the presence of
NP cannot be detected. We have shown that this type of
new physics can still be detected by performing an angular analysis.
In order to achieve the goal of visualising the effect of new 
physics in this mode, we first obtain six CP 
violating observables ($C_{1-6}$) from the angular distribution 
of decay products and show that within the SM, these observables 
are identically zero.
Any nonzero value found in the future study of these observables 
will indicate the presence of NP. Thereafter, we introduce the
generic new physics effect and obtain the modified $C$'s ( in the
presence of NP) and study in turn two beyond the SM scenarios for the sake
of illustration. In fact, we consider the VLDQ model and the RPV
supersymmetric model to look for NP.

In the VLDQ model we find that the first four ($C_{1-4}$) observables,
out of the six CP violating observables, are nonzero. 
If found so, this may indicate
the nature of the interaction Lagrangian in $B\to \phi^* K$ to be
of the form (V-A)(V-A), which is the case with the VLDQ model. Whereas
in the RPV model we find all the six observables to be nonzero. The 
nonzero values in terms of these observables will indicate the interaction
Lagrangian to be of (V-A)(V+A) form.

In summary, we studied the angular distribution analysis of the decay
$B\to \phi K^*$ in the SM and beyond it. We obtained six CP violating 
observables. These are vanishing in the SM but if found nonvanishing
in the future experiments, will definitely indicate the presence 
of new physics.
We have studied two promising models beyond the SM scenarios, where 
we have shown that these models indeed can have nozero $C$'s. 
Since no special technique is required to study them experimentally, and 
the data are already available, these findings can immediately be studied
to look for NP effects in $B\to \phi K^*$ in the currently running
$B$ factories. In fact, if these observables are found to be nonzero
experimentally then this in turn may eventually lead to the confirmation
of the (already existing speculation in $B\to \phi K_S$ decay) new physics
in the penguin dominated ($b\to s\bar s s$) $B$ decays.

To conclude, irrespective of the fact that whether NP is indeed present
in the $B\to \phi K^*$ decay mode or not, the study of the angular 
distribution will definitely rule out the possibility of the presence of
new physics or else establish a strong evidence of it. This angular
analysis study in turn deserves an immediate experimental attention.

\section{Acknowledgements}
The work of RM was supported in part by Department of Science and
Technology, Government of India through Grant No. SR/FTP/PS-50/2001.

\end{document}